\documentclass[12pt]{article}
\pdfoutput=1 

\usepackage[utf8]{inputenc}
\usepackage{jheppub}
\usepackage{epsf}
\usepackage{epsfig}
\usepackage{amsmath}
\usepackage{amsfonts}
\usepackage{amssymb}
\usepackage{graphicx}
\usepackage{longtable}
\usepackage{pdflscape}
\usepackage{subcaption}
\usepackage{color}
\usepackage{psfrag}
\usepackage{hyperref}
\usepackage{tikz}						
\usetikzlibrary{positioning,trees,decorations.pathmorphing,decorations.markings,decorations.pathreplacing,calc,shapes,patterns,arrows}
\usepackage[export]{adjustbox}
\usepackage{listings}
\usepackage{multirow} 
\usepackage{cancel,slashed}
\usepackage{bbm}
\usepackage{graphicx}
\usepackage{bbold}
\usepackage{latexsym}
\usepackage{color}
\usepackage{xypic}
\usepackage{transparent}
\usepackage{footmisc}
\usepackage{mathtools}
\usepackage{array}
\usepackage{makecell}
\usepackage{graphics,psfrag}
\usepackage{placeins}
\makeatletter
\def\@xfootnote[#1]{%
	\protected@xdef\@thefnmark{#1}%
	\@footnotemark\@footnotetext}
\makeatother

\newcommand{\bmat}{\left(\begin{array}}
	\newcommand{\emat}{\end{array}\right)}

\def\-{\hphantom{-}}

\def\s2{\frac{1}{\sqrt2}}

\def\beq{\begin{equation}}
\def\eeq{\end{equation}}
\def\beqa{\begin{eqnarray}}
\def\eeqa{\end{eqnarray}}

\def\tr{{\rm tr \,}}

\def\dim{{\rm dim \,}}

\def\Dsl{\,\raise.15ex\hbox{/}\mkern-13.5mu D} 

\DeclareMathOperator{\SU}{SU}
\DeclareMathOperator{\SO}{SO}

\DeclareMathOperator{\USp}{USp}

\newcommand{\CC}{\mathbb{C}}

\newcommand{\coma}{\text{ , }}
\newcommand{\fstop}{\text{ .}}

\newcommand{\UV}{\text{UV}}
\newcommand{\IR}{\text{IR}}

\makeatletter
\def\squarecorner#1{
    \pgf@x=\the\wd\pgfnodeparttextbox%
    \pgfmathsetlength\pgf@xc{\pgfkeysvalueof{/pgf/inner xsep}}%
    \advance\pgf@x by 2\pgf@xc%
    \pgfmathsetlength\pgf@xb{\pgfkeysvalueof{/pgf/minimum width}}%
    \ifdim\pgf@x<\pgf@xb%
        \pgf@x=\pgf@xb%
    \fi%
    \pgf@y=\ht\pgfnodeparttextbox%
    \advance\pgf@y by\dp\pgfnodeparttextbox%
    \pgfmathsetlength\pgf@yc{\pgfkeysvalueof{/pgf/inner ysep}}%
    \advance\pgf@y by 2\pgf@yc%
    \pgfmathsetlength\pgf@yb{\pgfkeysvalueof{/pgf/minimum height}}%
    \ifdim\pgf@y<\pgf@yb%
        \pgf@y=\pgf@yb%
    \fi%
    \ifdim\pgf@x<\pgf@y%
        \pgf@x=\pgf@y%
    \else
        \pgf@y=\pgf@x%
    \fi
    \pgf@x=#1.5\pgf@x%
    \advance\pgf@x by.5\wd\pgfnodeparttextbox%
    \pgfmathsetlength\pgf@xa{\pgfkeysvalueof{/pgf/outer xsep}}%
    \advance\pgf@x by#1\pgf@xa%
    \pgf@y=#1.5\pgf@y%
    \advance\pgf@y by-.5\dp\pgfnodeparttextbox%
    \advance\pgf@y by.5\ht\pgfnodeparttextbox%
    \pgfmathsetlength\pgf@ya{\pgfkeysvalueof{/pgf/outer ysep}}%
    \advance\pgf@y by#1\pgf@ya%
}
\makeatother

\pgfdeclareshape{square}{
    \savedanchor\northeast{\squarecorner{}}
    \savedanchor\southwest{\squarecorner{-}}

    \foreach \x in {east,west} \foreach \y in {north,mid,base,south} {
        \inheritanchor[from=rectangle]{\y\space\x}
    }
    \foreach \x in {east,west,north,mid,base,south,center,text} {
        \inheritanchor[from=rectangle]{\x}
    }
    \inheritanchorborder[from=rectangle]
    \inheritbackgroundpath[from=rectangle]
}

\hypersetup{
	pdftitle={No Go for a Flow},    
	pdfauthor={\textcopyright\ Federico Carta, Alessandro Mininno},     
	pdfsubject={HEP},   
	pdfcreator={pdfLaTex},   
	pdfproducer={LaTex}, 
	pdfkeywords={},
	colorlinks=false,
	linkcolor=blue,        
	citecolor=green,        
	filecolor=magenta,      
	urlcolor=blue, 
}

\begin{document}
	\pagestyle{plain}

	\makeatletter
	\@addtoreset{equation}{section}
	\makeatother
	\renewcommand{\theequation}{\thesection.\arabic{equation}}
	\pagestyle{empty}
\rightline{DESY 20-021}
\rightline{IFT-UAM/CSIC-20-24}
\vspace{3cm}

\begin{center}
	\LARGE{\bf No go for a flow}\\
	\large{Federico Carta\textsuperscript{1}, Alessandro Mininno\textsuperscript{2}\\[4mm]}
	\footnotesize{\textsuperscript{1}Deutches Electronen-Synchrotron, DESY,\\ Notkestra\ss e 85, 22607 Hamburg, Germany\\
		\textsuperscript{2}Instituto de F\'{\i}sica Te\'orica IFT-UAM/CSIC,\\[-0.3em] 
		C/ Nicol\'as Cabrera 13-15, 
		Campus de Cantoblanco, 28049 Madrid, Spain}\\
	\footnotesize{\href{mailto:federico.carta@desy.de}{federico.carta@desy.de}, \href{mailto:alessandro.mininno@uam.es}{alessandro.mininno@uam.es}}

	\vspace*{20mm}
	
	\small{\bf Abstract} \\
\end{center}
\begin{center}
	\begin{minipage}[h]{\textwidth}
		{\small We prove that a very large class of $15502$ general Argyres-Douglas theories cannot admit a UV lagrangian which flows to them via the Maruyoshi-Song supersymmetry enhancement mechanism. We do so by developing a computer program which brute-force lists, for any given 4d $\mathcal{N}=2$ superconformal theory $\mathcal{T}_{\text{IR}}$, all possible UV candidate superconformal lagrangians $\mathcal{T}_{\text{UV}}$ satisfying some necessary criteria for the supersymmetry enhancement to happen. We argue that this is enough evidence to conjecture that it is impossible, in general, to find new examples of Maruyoshi-Song lagrangians for generalized Argyres-Douglas theories. All lagrangians already known are, on the other hand, recovered and confirmed in our scan. Finally, we also develop another program to compute efficiently Coulomb branch spectrum, masses, couplings and central charges for $(G,G')$ Argyres-Douglas theories of arbitrarily high rank.}
	\end{minipage}
\end{center}

	\newpage
	\setcounter{page}{1}
	\pagestyle{plain}
	\renewcommand{\thefootnote}{\arabic{footnote}}
	\setcounter{footnote}{0}
	
	\tableofcontents
	

\section{Introduction}	

Four dimensional $\mathcal{N}=2$ quantum field theories received much interest in the past decades, as the large amount of supersymmetry allows one to perform exact computations even in the strongly coupled regime.

Soon after the discovery of Seiberg-Witten solutions \cite{Seiberg:1994aj,Seiberg:1994rs} it was realized that there exist consistent superconformal quantum field theories that do not admit a local lagrangian description, and are therefore named non-lagrangian theories \citep{Minahan:1996fg,Minahan:1996cj}. With the discovery of Argyres-Seiberg duality~\cite{Argyres:2007cn}, it was realized that such non-lagrangian theories are not just exotic sporadic examples of QFTs, but instead they are quite generic, and arise naturally as duals of ordinary lagrangian theories. Furthermore, the set of such non-lagrangian theories has been extremely extended with the class-S construction of Gaiotto~\cite{Gaiotto:2009we}.

In particular, one interesting set of strongly-coupled superconformal $\mathcal{N}=2$ non-lagrangian theories are the so called Argyres-Douglas (AD) theories. The defining property of an Argyres-Douglas theory is that it exists at least one Coulomb Branch (CB) operator that has a fractional (non-integer) conformal dimension. Argyres-Douglas theories were originally found to describe the low-energy dynamics at special point in the Coulomb Branch moduli space of a pure $\mathcal{N}=2$ super Yang-Mills with simply-laced gauge group $G$, where at the same time mutually non-local dyons become massless~\cite{Argyres:1995jj, Argyres:1995xn}. In the following we will denote Argyres-Douglas theories of this type as $G$-Argyres-Douglas, or equivalently $(A_1, G)$ theories.

In~\cite{Cecotti:2010fi}, this class of AD theories has been enlarged. It was shown that by compactifying type IIB superstring theory on a Calabi-Yau 3-fold singularity given by the sum of two $ADE$ polynomials, one could recover the $(A_1, G)$ theories, as well as construct many more. The resulting $4$d $\mathcal{N}=2$ superconformal theories obtained in such way are called $(G,G')$ theories, where $G$ and $G'$ are the two $ADE$ type groups that define the CY singularity. Equivalently, $(G,G')$ theories could also be defined by the fact that their BPS quiver~\cite{Alim:2011kw} is the direct product of two Dynkin diagrams of type $G$ and $G'$. It was shown that a subset of these theories, precisely those of the form $(A_n,G)$ admit a class-S description. The Riemann surface is a sphere, and there is a single irregular puncture on it \cite{Bonelli:2011aa, Xie:2012hs, Wang:2015mra}. This was further generalized to the case of twisted punctures in \cite{Wang:2018gvb}.

Recently it was remarkably found that the set of non-lagrangian $4$d $\mathcal{N}=2$ theories is somehow smaller than what initially thought. Indeed, some $\mathcal{N}=1$ Lagrangian gauge theories were found by Maruyoshi and Song (MS) to flow to some of the $(G,G')$ in the deep IR, therefore showing a phenomenon of Supersymmetry Enhancement at low energies \cite{Maruyoshi:2016aim,Maruyoshi:2016tqk}. For a complementary approach, see~\cite{Agarwal:2018ejn,Buican:2018ddk,Gadde:2015xta,Maruyoshi:2018nod, Apruzzi:2018xkw,Razamat:2019vfd}.

The idea of Maruyoshi and Song (MS) in~\cite{Maruyoshi:2016aim,Maruyoshi:2016tqk} was to consider a $\mathcal{N}=2$ superconformal field theory $\mathcal{T}_{\UV}$ with a non-abelian flavor symmmetry $F$, and to deform it by adding a superpotential term in which a gauge-singlet, flavor adjoint $\mathcal{N}=1$ chiral multiplet $M$ couples to the moment map operator $\mu$ via a superpotential term, 
\begin{equation}
    W_{\text{def}}=\tr M \mu\fstop
\end{equation}
One then gives a nilpotent vacuum expectation value (vev) $\langle M\rangle$ to $M$, therefore triggering an RG flow. Depending on the choice of $\mathcal{T}_{\UV}$ and $\langle M\rangle$, it is found that the IR theory $\mathcal{T}_{\IR}$ could be $\mathcal{N}=2$ again, and if it is so, then $\mathcal{T}_{\IR}$ is often one of the $(G,G')$ theories. 

Such proposal was checked in two different ways. In the first one, the superconformal central charges $(a,c)$ of $\mathcal{T}_{\IR}$ are recovered by the a-maximization technique \citep{Intriligator:2003jj}. In the second, the full superconformal index \citep{Kinney:2005ej} was computed for $\mathcal{T}_{\UV}$ and it was shown that its Schur limit, Macdonald limit, and Coulomb limit all agree with the ones of $\mathcal{T}_{\IR}$.

Let us give an example of such flows. Consider $\mathcal{T}_{\UV}$ to be $\SU(2)$ with $N_f=4$. The flavor symmetry is $\SO(8)$, and one can give to $M$ a vev inside the maximal nilpotent orbit of the Lie algebra $\mathfrak{so}(8)_{\mathbb{C}}$. The resulting IR theory is the $(A_1,A_2)$ Argyres-Douglas theory, also known as $H_0$, the minimal 4d $\mathcal{N}=2$ SCFT\footnote{Here we mean that the $(A_1, A_2)$ theory has the minimal known value of central charges $a$ and $c$. For the central charge $c$, it is proven both by Bootstrap argument and chiral algebras \cite{Liendo:2015ofa, Cornagliotto:2017snu} that it is impossible to find a $\mathcal{N}=2$ SCFT with a lower $c$ than the one of the $(A_1, A_2)$ theory.}.

The Maruyoshi-Song deformation was later generalized to the case in which the field $M$ acquires a vev along some non-maximal nilpotent orbit, in~\cite{Agarwal:2016pjo}. In~\cite{Agarwal:2017roi, Benvenuti:2017bpg} it was also applied in cases in which the UV theory $\mathcal{T}_{\UV}$ is a linear superconformal quiver. At the moment, it is known that the following set of $(G,G')$ theories admit a MS lagrangian:
\begin{align}
    &\text{\textbullet \hspace{1mm} $\left(A_{m-1},A_{Nm-1}\right)$ theory with $m, N\geq 1$.}\nonumber\\
    &\text{\textbullet \hspace{1mm} $\left(A_{2m-1},D_{2Nm+1}\right)$ theory with $m, N\geq 1$.} \label{list:knownlag}\\
    &\text{\textbullet \hspace{1mm} $\left(A_{2m},D_{m(N-2)+\frac{N}{2}}\right)$ theory with $m\geq 1$ and $N\geq 4$ and even.}\nonumber
\end{align}
After their first introduction, Maruyoshi-Song RG-flows were further studied. The compactification to 3 dimensions was studied in \citep{Benvenuti:2017lle,Benvenuti:2017kud,Benvenuti:2017bpg,Agarwal:2018oxb}. In \citep{Giacomelli:2017ckh} it was observed that all known MS flows admit a class-S description in which $\mathcal{T}_{\UV}$ has a class-S realization as a sphere $S^2_{\UV}$ with one irregular and one regular full puncture and $\mathcal{T}_{\IR}$ has a class-S realization as a sphere $S^2_{\IR}$ with one irregular puncture alone. The order of the pole of the Hitchin field at the irregular puncture of $S^2_{\IR}$ is increased by one, with respect of the order of the pole of the Hitchin field at the puncture of $S^2_{\UV}$. In \citep{Carta:2019hbi} the enhancement phenomenon is studied at the level of the Hitchin system. Furthermore, necessary criteria to establish if a theory can admit supersymmetry enhancement using such deformation have been introduced by Giacomelli in~\cite{Giacomelli:2018ziv}, by exploiting 't Hooft anomaly matching conditions. In \citep{Carta:2018qke} a F-theory embedding of MS flows among rank 1 theories was presented.

In this paper we address the question whether the set (\ref{list:knownlag}) of $(G,G')$ theories $\mathcal{T}_{\IR}$ for which a Maruyoshi-Song UV lagrangian exists is complete, or we could maybe find lagrangians $\mathcal{T}_{\UV}$ for other $(G,G')$ theories. 

We show the results of a scan done over 20100 $(G,G')$ theories. We wrote a computer program that lists all the possible UV theories satisfying Giacomelli's necessary criteria for SUSY enhancement to happen, given as an input the theory $\mathcal{T}_{\IR}$. We stress that our computer program does not rely on the hypothesis that the IR theory is of $(G,G')$ type: the program is completely general. Given as an input the IR theory $\mathcal{T}_{\IR}$, the program will give as an output all its possible MS UV completions. It was simply our choice to look for candidate lagrangians for $(G,G')$ theories and not some other set of theories as, for example, class-S with regular punctures. We also stress that if the program gives a negative result this implies such lagrangian does not exist.

The result of our scan is as follows. First of all, we decided to abort the computation for any $(G,G')$ theory for which coming to a definitive answer took more than 6 hours of computing time on a 16 cores machine. We chose this 6-hours mark as the best compromise between giving the program sufficient time to work on each case, and being able to complete the full scan in a timescale of 3 months.

Out of the 20100 cases we chose to focus on, for 15999 of them the algorithm terminated in less than six hours of computing time. For 15502 out of the 15999 completely analyzed cases, a Maruyoshi-Song UV lagrangian has been proven \textbf{not} to exist. For the remaining 497 analyzed cases, a lagrangian already known in the literature was recovered (they are cases in the list of Eq.~\eqref{list:knownlag}). For the 4101 cases in which the algorithm did not terminate in less than 6 hours, our program could not come up with a definitive answer. For those latter cases we still cannot claim that a MS UV lagrangian does not exist.

This negative result made us conjecture that for all the AD theories of the $(G,G')$ type, all the Maruyoshi-Song lagrangians that flow to them are already known. This is also consistent with the conjecture that a Maruyoshi-Song flow between $\mathcal{T}_{\UV}$ and $\mathcal{T}_{\IR}$ exists if and only if they admit the class-S realization with the punctured spheres, as discussed above. Our scan gives strong explicit evidence in support of the validity of this latter statement.

This paper is organized as follows. In Section~\ref{sec:Argyres-Douglas} we review how $(G,G')$ theories are described from IIB geometrical engineering, and we introduce a computer program to compute their CB spectrum and central charges in a very efficient way. In Section~\ref{sec:UVcompletions} we first review an algorithm to check for UV lagrangian theories which flows to the a given $\mathcal{T}_{\IR}$ theories via the Maruyoshi-Song deformation. Then we introduce another computer program that implements such algorithm efficiently. In Section~\ref{sec:results} we describe the results of the scan we have done looking for new examples of UV lagrangians in the $(G,G')$ landscape. We state properly a well-motivated conjecture about the non-existence of them.

Both computer programs as well as a guide explaining the details of the code are publicly available as ancillary files.

\section{Geometrical engineering for $(G,G')$ theories}
\label{sec:Argyres-Douglas}

In this section we will recall how to realize the $(G,G')$ theories from IIB geometrical engineering \cite{Cecotti:2010fi}. Such realization of the field theory is particularly useful to compute the spectrum of Coulomb Branch operators, masses, couplings and central charges in a simple and algorithmic way.

Consider type IIB Superstring Theory compactified on $\mathbb{R}^{1,3}\times X$, where $X$ is a non-compact Calabi-Yau 3-fold singularity given by the zero-locus of the equation
\begin{equation}
W(x,y,z,w):=W_G(x,y)+W_{G'}(z,w)=0\coma
\label{eq:pGG'}
\end{equation}
where $(x,y,z,w)\in\CC^4$, while $W_G$ and $W_{G'}$ are the ADE polynomials:
\begin{equation}
\begin{split}
W_{A_n}(x,y)&=x^{n+1}+y^2\coma\\
W_{D_n}(x,y)&=x^{n-1}+xy^2\coma\\
W_{E_6}(x,y)&=x^3+y^4\coma\\
W_{E_7}(x,y)&=x^3+xy^3\coma\\
W_{E_8}(x,y)&=x^3+y^5\fstop
\end{split}
\label{eq:pADE}
\end{equation} 

We define the ring polynomials of four complex variables $\CC[x,y,z,w]$ modded by the ideal generated by the gradient $dW$~\cite{Shapere:1999xr},
\begin{equation}
\mathcal{R}=\CC[x,y,z,w]/dW\fstop
\end{equation}
Let us call $x^\alpha\in \mathcal{R}$ the monomials that generate $\mathcal{R}$. Each one of such monomials defines a deformation of the Calabi-Yau \eqref{eq:pGG'} of the form
\begin{equation}
W(x,y,z,w)\longrightarrow W(x,y,z,w)+\sum_{x^\alpha\in\mathcal{R}}u_\alpha x^\alpha\coma
\end{equation}
where the coefficients $u_{\alpha}$ will be interpreted as CB operators, masses or coupling constants depending whether their scaling dimension is respectively greater than one, equal to one\footnote{There can be masses with dimensions greater than $1$ but they are not paired up with other parameters such that their dimensions sum up to $2$.}, or smaller than one.

The scaling dimension of the parameters can be computed as follows. On the Calabi-Yau given by~\eqref{eq:pGG'}, it is naturally defined an holomorphic $3-$form $\Omega$, which locally reads
\begin{equation}
\Omega=\frac{dx\wedge dy \wedge dz \wedge dw}{dW}\fstop
\label{eq:Omega}
\end{equation} 
Such holomorphic 3-form has scaling dimension $1$, since BPS masses can be computed as periods of~\eqref{eq:Omega} on supersymmetric $3-$cycles~\cite{Shapere:1999xr}. The condition $[\Omega]=1$, together with the homogeneity of Eq.~\eqref{eq:pGG'} allows us to solve for all the scaling dimensions.

Given the spectrum of the CB operators, it is then possible to compute also the superconformal central charges $(a,c)$ of the field theory using the following relations~\cite{Shapere:2008zf}:
\begin{equation}
a=\frac{1}{4}R(A)+\frac{1}{6}R(B)+\frac{5}{24}r \coma \quad  c=\frac{1}{3}R(B)+\frac{1}{6}r\coma
\end{equation}
where 
\begin{equation}
R(A)=\sum_{i}[u_i]-r\fstop
\end{equation}
and $R(B)$ is related to the discriminant of the Seiberg-Witten curve. For the particular case of the class of theories of interest, $R(B)$ can be easily computed as~\cite{Cecotti:2013lda,Cecotti:2015lab} 
\begin{equation}
R(B)=\frac{r_{G}r_{G'}}{4}\frac{h_{G}^{\vee}h_{G'}^{\vee} }{h_{G}^{\vee}+h_{G'}^{\vee}}\coma
\end{equation} 
where $h_{G}^{\vee}$ is the dual Coxeter number of the group $G$, and $r_{G}$ is the rank of $G$.

\subsection{A program to compute central charges}
\label{sec:RACprogra}

The computation described in Section~\ref{sec:Argyres-Douglas} may result tedious when the rank of groups becomes large. Attached to this paper there is a Mathematica notebook, called \texttt{GGp\_RAC.nb} that does the computation for us. The program is quite straightforward to understand since it applies literally the computation described in Section~\ref{sec:Argyres-Douglas}. However, in the ancillary files there is a ``Guide\_programs" that explains in details what are the necessary inputs for the program to work efficiently. Since we use the Type IIB description for $(G,G')$ theories, the function needs as input two of the following semisimple Lie algebras
\begin{equation}
A_{n\geq 1} \coma D_{n\geq 3} \coma E_{6} \coma E_7 \coma E_8\coma
\end{equation}
and it returns as output an array containing:
\begin{itemize}
	\item The scaling dimensions of the Coulomb Branch operators.
	\item The number of masses, i.e. the rank of the flavor symmetry group.
	\item The complex dimension of the Coulomb Branch.
	\item The superconformal central charges $a$ and $c$.
	\item The complex dimension of the Higgs Branch as\footnote{In some cases this formula may give fractional results. In those cases, there is no Higgs branch.}
	\begin{equation}
	    \dim \text{HB}=24(c-a)\fstop
	\end{equation}
\end{itemize}
The notebook is set to work on Linux distributions. The program will launch a Macaulay2~\cite{M2} subroutine, which is used to compute the ideal of the gradient of~\eqref{eq:pGG'}, therefore it is necessary to pre-install Macaulay2.

\begin{figure}[h!]
\centering
    \begin{minipage}{\textwidth}
    \centering
    \includegraphics[width=\textwidth,keepaspectratio]{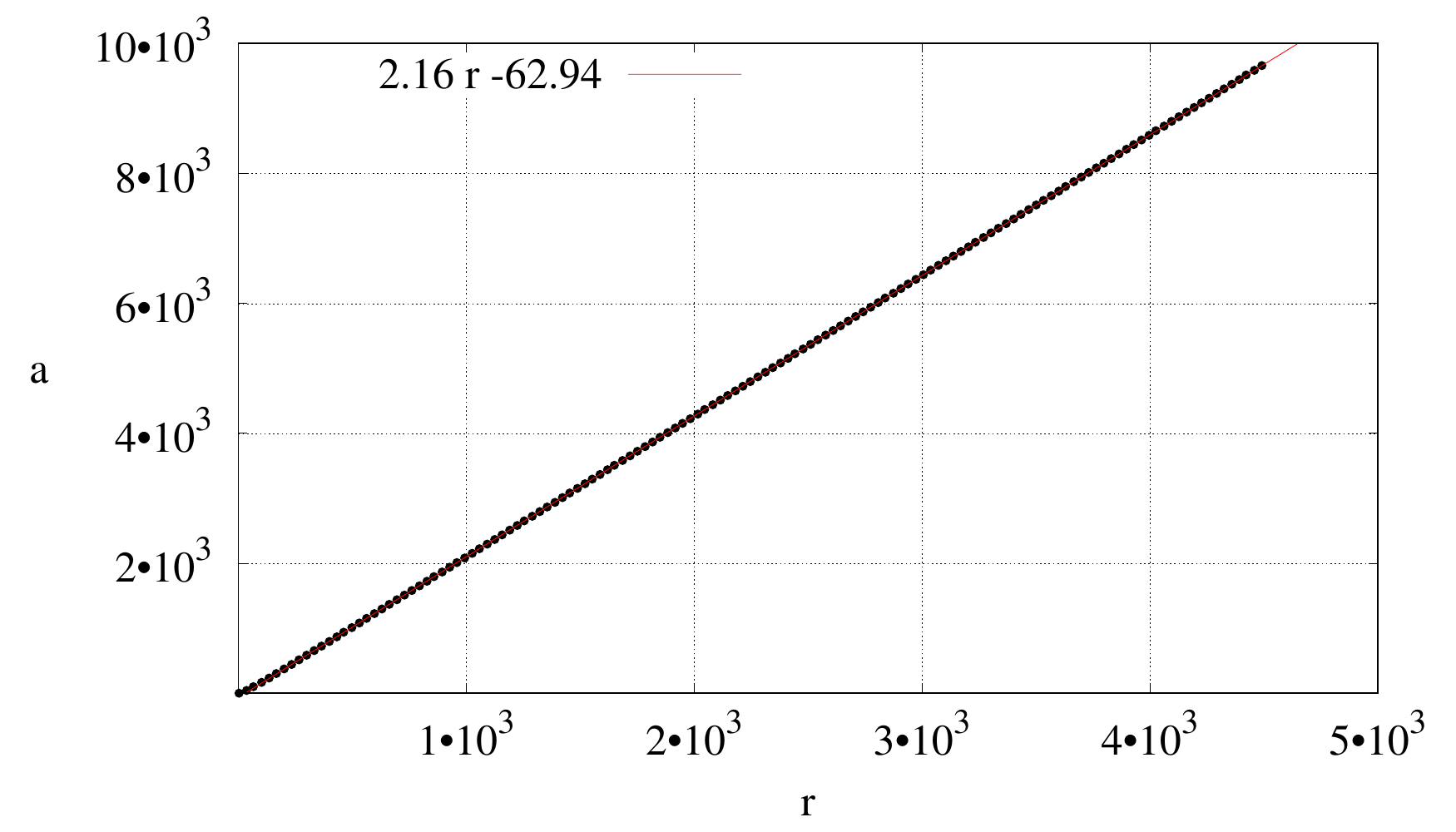}
    \end{minipage}
    \begin{minipage}{\textwidth}
    \centering
    \includegraphics[width=\textwidth,keepaspectratio]{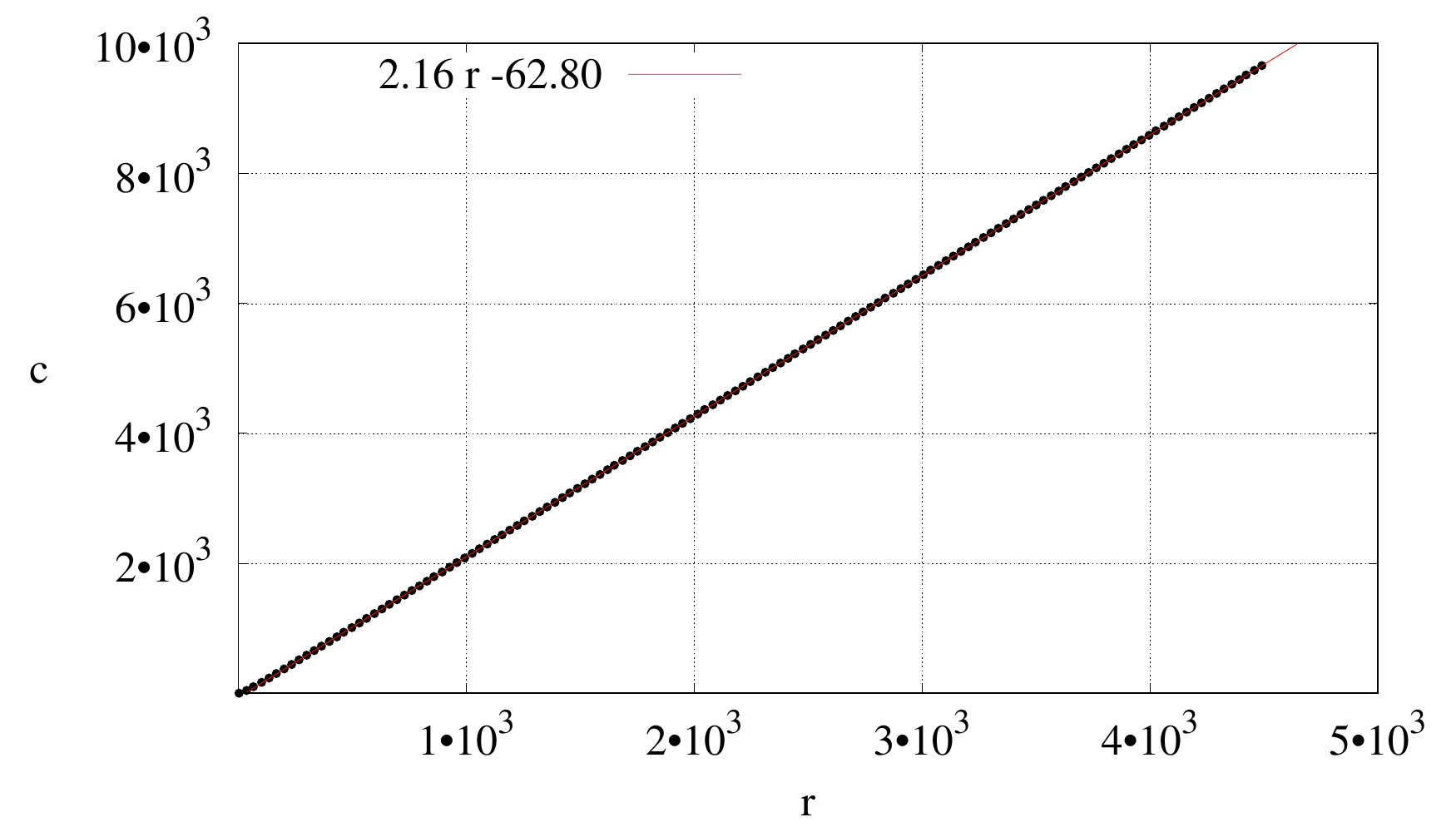}
    \end{minipage}
    \caption{We plot how the central charges scale with the rank for $(A_m,E_6)$ with $m$ in $[1,1500]$.}
    \label{fig:E6A1500rac}
\end{figure}

As an immediate application of such program we can easily compute the central charges for $(G,G')$ theories of very large ranks. For example, we can check the fact already noticed in~\cite{Xie:2013jc} that for $(G,G')$ theories the central charges $a$ and $c$ scale linearly with the rank. Figure~\ref{fig:E6A1500rac} shows the case of $(A_{m},E_6)$ for all $m$ in $[1,1500]$. 

\section{An algorithm to look for candidate UV completions}
\label{sec:UVcompletions}

\subsection{The algorithm}

Consider a $4$d $\mathcal{N}=2$ superconformal field theory $\mathcal{T}_{\IR}$. In this Section we are going to review an algorithm that allows us either to find all the $4$d $\mathcal{N}=2$ lagrangian SCFTs $\mathcal{T}_{\UV}$ that flow to $\mathcal{T}_{\IR}$ under a MS deformation, or to prove that such a lagrangian UV completion for $\mathcal{T}_{\IR}$ cannot exist.

Such algorithm was originally introduced in \cite{Giacomelli:2018ziv}, where some necessary conditions for the existence of $\mathcal{T}_{\UV}$ were found via an argument of 't Hooft anomaly matching~\cite{tHooft:1979rat} for the R-symmetries of $\mathcal{T}_{\IR}$ and $\mathcal{T}_{\UV}$. Here we will need some of these conditions, namely

\begin{enumerate}
\item The rank  $r$ of $\mathcal{T}_{\IR}$ is equal to the rank of $\mathcal{T}_{\UV}$, namely $r_{\UV}=r_{\IR}=r$.
\item The central charges of $\mathcal{T}_{\UV}$ and $\mathcal{T}_{\IR}$ are related as follows
\begin{equation}
(6c_{\IR}-r)(4a_{\UV}-5c_{\UV})=(6c_{\UV}-r)(4a_{\IR}-5c_{\IR})\fstop
\label{eq:acUVacIRrel}
\end{equation}
\item The number of simple factors $f$ of the gauge group of $\mathcal{T}_{\UV}$ is equal to the number of CB operators of $\mathcal{T}_{\IR}$ of smallest conformal dimension.
\end{enumerate}

Recall now that we focus only on the case in which $\mathcal{T}_{\UV}$ is lagrangian. Crucially then Eq.~\eqref{eq:acUVacIRrel} can be rewritten in terms of the number of hypermultiplets $n_h$ and vector multiplets $n_v$ of $\mathcal{T}_{\UV}$ using the usual formulae for weakly coupled theories,
\begin{equation}
a_{\UV}=\frac{4n_v+n_h}{24} \coma \ c_{\UV}=\frac{2n_v+n_h}{12}\fstop
\end{equation} 

We find then that the number of hypermultiplets of $\mathcal{T}_{\UV}$ is given by

\begin{equation}
n_h=4\frac{(4 a_{{\IR}}-5c_{\IR}) (r-n_v)}{8a_{\IR}-4 c_{\IR}-r}\fstop
\label{eq:nh}
\end{equation}

In the following we assume we have knowledge of $f$, $r$, $a_{\IR}$ and $c_{\IR}$ for our given theory $\mathcal{T}_{\IR}$. Now the algorithm proceeds as follows. 

\begin{itemize}
    \item We plug into \eqref{eq:nh} the IR central charges and the rank. Then, we list all the possible gauge groups of the UV theory having exactly $f$ simple factors and having rank $r$. Clearly, there is a finite number of them. For each such choice of $G_{\UV}$, the number of vector multiplets $n_v$ is equal to the dimension of $G_{\UV}$, then we can solve \eqref{eq:nh} for $n_h$.

\item If by this computation we find a non-integer value for $n_h$, we conclude that a lagrangian theory $\mathcal{T}_{\UV}$ which flows to $\mathcal{T}_{\IR}$ under a MS deformation cannot exist.

\item If instead we find a integer value for $n_h$, the algorithm continues as follows. We list all the possible gauge theories that can be formed by using the selected gauge group $G_{\UV}$ and the number of hypermultiplets such determined. In particular, we will need to split the $n_h$ ``loose hypermultiplets" into representations of the various $f$ factors of $G_{\UV}$. This number is clearly finite as both $n_h$ and $f$ are.

\item Out of all these possible ways of assigning the hypers to some gauge representation, we compute whether the beta-function for all the factors of the gauge group $G_{\UV}$ vanishes. If not, we drop such case. This drastically reduces the possibilities. This last check is based on the classification of lagrangian $\mathcal{N}=2$ SCFTs made in~\cite{Bhardwaj:2013qia}. 

\item If a non-trivial way to assign the $n_h$ hypers into representations of $G_{\UV}$ such that all beta functions vanish is found, then such theory could be a UV completion for $\mathcal{T}_{\IR}$. However, such possibility can be still excluded, for instance by checking whether it is free of Witten's anomaly~\cite{Witten:1982fp} or checking by $a-$maximization~\cite{Intriligator:2003jj} whether it really flows to $\mathcal{T}_{\IR}$.
\end{itemize}

We stress that this algorithm crucially relies on the assumption that $\mathcal{T}_{\UV}$ is lagrangian. Even when algorithm rules out a Maruyoshi-Song lagrangian UV completion of a given $\mathcal{T}_{\IR}$ theory, it is still possible (and in fact it happens in various examples) that $\mathcal{T}_{\IR}$ can admit a Maruyoshi-Song non-lagrangian UV completion.

\subsection{The implementation}

In this section we discuss the main features of a computer program we realized in order to perform the analysis of Section~\ref{sec:UVcompletions}. This program is available as an ancillary file \texttt{UVtheory.nb}, together with a file ``Guide\_programs" containing a more detailed documentation about how the code works.

Given as an input the rank $r_{\IR}$, the central charges $a_{\IR}$ and $c_{\IR}$ and the number $f$ of CB operators with the smallest dimension of any given 4d $\mathcal{N}=2$ SCFT $\mathcal{T}_{\IR}$, the function \texttt{UVTheory} computes all the possible UV lagrangian theories which could flow to $\mathcal{T}_{\IR}$ under a Maruyoshi-Song deformation. The code is then completely general because it needs only information of the IR theory, $\mathcal{T}_{\IR}$ and it will provide an output with all the candidates UV theories as follows.

The function \texttt{UVTheory} will first compute all the possible choices of $f$ simple groups whose total rank is equal to $r$. From the dimension of the simple groups, it computes the number $n_v$ of vector multiplets of $\mathcal{T}_{UV}$ and using~\eqref{eq:nh} it computes the number of loose hypermultiplets. For every single (resp. couple, or triplet) of factors among the $f$ simple groups the possibility of having an hyper charged under it (resp under both of them, or the three of them) is considered, checking that such choice is compatible with the classification of possible $n-$gons given by~\cite{Bhardwaj:2013qia}. 
In more details, the program sorts in all possible ways the loose hypermultiplets into all possible allowed representations listed in Tables 1, 2 and 3 of~\cite{Bhardwaj:2013qia}. Then for each of these possibilities the beta-function contribution for every factor is computed, and non-conformal cases are dropped.

The program will then give as an output the list of UV theories which pass these criteria. Let us see at one concrete example in details.

 Consider the $(A_3, A_7)$ Argyres-Douglas theory. This theory has
\begin{equation}
r=9 \coma a = \frac{145}{24} \text{ and } c = \frac{37}{6}\coma
\end{equation}
and the number of CB operators with smallest dimension is
\begin{equation}
    f=3\fstop
\end{equation}
\begin{figure}[h!]
    \centering
    \begin{tikzpicture}[scale=1]
\node[draw,circle] (A) at (180:4.5) {$\SU(2)$};
\node[draw,circle] (B) at (180:1.5) {$\SU(4)$};
\node[draw,circle] (C) at (0:1.5) {$\SU(6)$};
\node[draw,square] (D) at (0:4.5) {$\SU(8)$};
\draw[line width=1pt] (A) to (B);
\draw[line width=1pt] (B) to (C);
\draw[line width=1pt] (C) to (D);
    \end{tikzpicture}
    \caption{Quiver for the theory UV completion of $(A_3,A_7)$.}
    \label{fig:A1A3A5quiver}
\end{figure}
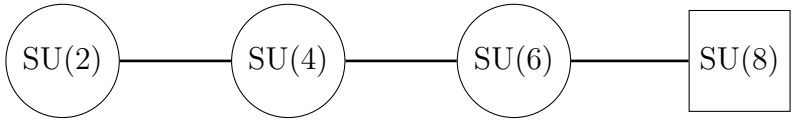
We know from~\cite{Agarwal:2017roi, Benvenuti:2017bpg}, that this theory has a UV completion in the quiver in Figure~\ref{fig:A1A3A5quiver}, however, we want to show how to read such theory from the output of the function \texttt{UVTheory} of our program. The output of \texttt{UVtheory} will be the following: 
\begingroup
\allowdisplaybreaks
\begin{subequations}
\begin{align}
\begin{split}
\{\\
\{\\
& \left\{\text{AA}_1 \text{AA}_3 \text{AA}_5,53,80\right\},\\
&\{\},\\
&\{\left\{\text{AA}_1,\text{AA}_3\right\},\left\{\text{AA}_3,\text{AA}_5\right\}\},\\
&\{\left\{\text{AA}_1\right\},\left\{\text{AA}_3\right\},\left\{\text{AA}_5\right\}\},\\
\{\{&y(1,1,1)\to 1,y(2,1,1)\to 1,\\
&z(1,1)\to 0,z(1,3)\to 0,z(1,4)\to 0,\\
&z(2,1)\to 0,z(2,3)\to 0,z(2,4)\to 0,\\
&z(3,1)\to 8,z(3,2)\to 0,z(3,3)\to 0,z(3,4)\to 0\}\}\\
\}\\
\end{split}\label{eq:exampleoutput1}\\
\begin{split}
\{\\
& \left\{\text{AA}_4 \text{DD}_1 \text{DD}_4,53,80\right\},\\
&\{\},\\
&\{\left\{\text{AA}_4,\text{DD}_4\right\}\},\\
&\{\left\{\text{AA}_4\right\},\left\{\text{DD}_1\right\},\left\{\text{DD}_4\right\}\},\\
\{\{&y(1,1,1)\to 1,\\
&z(1,1)\to 2,z(1,2)\to 0,z(1,3)\to 0,z(1,4)\to 0,\\
&z(2,1)\to 0,z(2,4)\to 22,\\
&z(3,1)\to 1,z(3,4)\to 0\}\}\\
\}\\
\end{split}\label{eq:exampleoutput2}\\
\begin{split}
\{\\
& \left\{\text{AA}_5 \text{CC}_1 \text{DD}_3,53,80\right\},\\
&\{\},\\
&\{\left\{\text{CC}_1,\text{DD}_3\right\}\},\\
&\{\left\{\text{AA}_5\right\},\left\{\text{CC}_1\right\},\left\{\text{DD}_3\right\}\},\\
\{\{&y(1,1,1)\to 1,\\
&z(1,1)\to 4,z(1,2)\to 2,z(1,3)\to 0,z(1,4)\to 0,\\
&z(2,1)\to 2,z(2,4)\to 0,\\
&z(3,1)\to 3,z(3,4)\to 0\}\}\\
\}\\
\}\fstop
\end{split}\label{eq:exampleoutput3}
\end{align}
\label{eq:exampleoutput}  
\end{subequations}
\endgroup

The interpretation of the output is as follows. First of all, in general, there can be many combinations of $f$ groups such that their rank is $r$. The output is then an array containing all possible combinations that are allowed by~\cite{Bhardwaj:2013qia}. In this case, in fact, there are in principal three different combinations of simple groups:
\begin{itemize}
    \item $G=A_1\times A_3 \times A_5$ in Eq.~\eqref{eq:exampleoutput1};
    \item $G=A_4\times D_1\times D_4$ in Eq.~\eqref{eq:exampleoutput2};
    \item $G=A_5\times C_1\times D_3$ in Eq.~\eqref{eq:exampleoutput3}.
\end{itemize}
The first component of each output, then, always contains the gauge group of the UV theory, followed by the number of vector multiplets and the number of loose hypermultiplets. They, in this case, are, respectively,
\begin{equation}
    n_v=53 \text{ and } n_h=80 \fstop
\end{equation}
The following component contains the combinations of triple factors under which a field can be in the trifundamental of all of them. In the example at hand, there is not allowed triple of factors of gauge groups such that it can admit trifundamentals. However, there can be couple of groups under which a field can be charged at the same time. The allowed couples are in the following component of the array. For instance, in Eq.~\eqref{eq:exampleoutput1}, there are two possible couples: $A_1\times A_3$ and $A_3\times A_5$. Under such groups, there can be hypermultiplets in some representation. It is worth to notice that Eq.~\eqref{eq:exampleoutput2} and Eq.~\eqref{eq:exampleoutput3} contain allowed couples, but not all the factors belong to a couple. This means that the theory will be formed by a disconnected quiver. In our analysis we drop such cases by hand. 

The last component involving the groups lists all the allowed single group factors that the UV theory admits. We are now almost ready to read the last component of the Eq.~\eqref{eq:exampleoutput}. This component contains a way to distribute the loose hypermultiplets among the allowed triplets, pairs or single factors. The variables $y$ and $z$ are associated respectively to pairs or single factors of the gauge groups\footnote{In case in which a triplet is allowed, the program associate to it a variable called $x$.}. Let us consider a generic $y(i,j,k)$ variable as example. The $i-$component is the number of the pair that is referring to. For Eq.~\eqref{eq:exampleoutput1}, $y(1,j,k)$ corresponds to the couple $\{A_1,A_3\}$. The components $j$ and $k$ are associated to the representation that the hypermultiplet has under respectively the first and the second element of the pair. 

This is a computational trick that creates a dictionary between the component of the array and the corresponding representation. In other contexts, such trick can be thought as an hash table. It works as follows. In~\cite{Bhardwaj:2013qia} for all possible simple groups there are $12$ types of possible representations\footnote{It is important to stress that not all the groups allow for all the $12$ types of representations, but if someone lists all the different representations that are allowed for all the groups, they are $12$.}. They are

\begin{enumerate}
	\item Fundamental / Vector representation for $\SU(n)$, $\SO(n)$ and $\USp(n)$.
	\item Antisymmetric representation for $\SU(4)$ and $\USp(n)$.
	\item Symmetric representation for $\SU(n)$.
	\item Adjoint representation for $\SU(n)$, $\SO(n)$, $\USp(n)$, $\text{E}_6$, $\text{E}_7$, $\text{E}_8$, $\text{F}_4$ and $\text{G}_2$.
	\item 3-index antisymmetric representation for $\SU(6)$, $\SU(7)$, $\SU(8)$, $\USp(6)$ and $\USp(8)$.
	\item S - spinorial representation of $\SO(n)$ for $7\leq n \leq 14$.
	\item C - conjugate spinorial representation of $\SO(8)$ and $\SO(12)$.
	\item 16 dimensional representation of $\USp(4)$.
	\item 27 dimensional representation of $\text{E}_6$.
	\item 56 dimensional representation of $\text{E}_7$.
	\item 26 dimensional representation of $\text{F}_4$.
	\item 7 dimensional representation of $\text{G}_2$.
\end{enumerate}

Each component $j$ and $k$ of $y(i,j,k)$ goes from $1$ to $12$ telling us what is the representation of the hypermultiplet under the group. A concrete example can, again, be done using $y(1,1,1)$ of Eq.~\eqref{eq:exampleoutput1}. We have said that this element is associated to the pair $\{A_1,A_3\}$ and it represents an hypermultiplet in the bifundamental representation of these groups. From the list of possible representation, it is clear that not all the groups can admit such representation, because some of them are specific for some particular case, as explained in the tables in~\cite{Bhardwaj:2013qia}.

The elements $z(n,m)$ work in the same way: in this case we are looking at the $n-$th element of the list of single factors in the representation corresponding to the letter $m$. For an extravagant example, in Eq.~\eqref{eq:exampleoutput3} there are $2$ hypermultiplets in the antisymmetric representation of $\SU(6)$. However, as said before, Eq.~\eqref{eq:exampleoutput2} and~\eqref{eq:exampleoutput3} are not corresponding to connected quiver, so we are not interested in them.\\
If a triplet had had been allowed in this example, the corresponding hypermultiplet will be associated to the variable $x(1,j,k,l)$, with $j$, $k$, $l$ representing its representation under the three gauge groups.\\ 

Now that we have understood how to read the output of \texttt{UVTheory}, it should be easy to see that Eq.~\eqref{eq:exampleoutput1} corresponds to the quiver in Figure~\ref{fig:A1A3A5quiver}. We have, then, obtained the UV theory found also in~\cite{Agarwal:2017roi, Benvenuti:2017bpg} for the $(A_3,A_7)$ Argyres-Douglas theory. In a similar fashion, our program reproduces all known results of UV lagrangians that flow to AD theories of the type $(G,G')$, but, since the input are very general, it may be useful to test new or more complicated theories.

\section{Results}
\label{sec:results}

In~\cite{Agarwal:2017roi,Maruyoshi:2016aim,Maruyoshi:2016tqk,Agarwal:2016pjo,Benvenuti:2017bpg} many UV lagrangians that flow to AD theories of the type $(G,G')$ have been found. We wanted to use our programs to test if there are some other UV lagrangians to be found for theories in the $(G,G')$ landscape. We ran the program for the following sample of 20100 theories:
\begin{equation}
\begin{split}
    &(A_n,A_m) \coma (A_n,D_m) \coma (D_n,D_m) \coma\\
    &(A_n,E_6) \coma (A_n,E_7) \coma (A_n,E_8) \coma\\
    &(D_m,E_6) \coma (D_m,E_7) \coma (D_m,E_8)\coma\\
    &(E_6,E_6) \coma (E_6,E_7) \coma (E_6,E_8) \coma \\
    &(E_7,E_7) \coma (E_7,E_8) \text{ and } (E_8,E_8)\coma 
\end{split}
\label{casesanalized}
\end{equation}
with $n=1,\ldots, 100$ and $m=3,\ldots,100$. 

One difficulty of such scan is the time that the program needs to find all possible candidates UV theories that have a fixed rank $r$ and a product of $f$ simple groups, before testing for the vanishing beta-function. The number of such combinations scales exponentially both in $r$ and $f$, so we decided to interrupt the computation for each $(G,G')$ theory if after six hours it was not terminated. The main result that we find is the following:\\

\noindent\fbox{%
    \parbox{\textwidth}{
\textit{\noindent For all cases in (\ref{casesanalized}) for which a Maruyoshi-Song UV lagrangian is not already know in the literature, and for which an output of our program was produced within 6 hours, we find that such UV lagrangian cannot exist.
}
}%
}\\

We have decided to put the results of the scan in tables from~\ref{tab:AnAm1} to~\ref{tab:EnEm} for an immediate and easy reading. The tables contain green, red or gray boxes. If the program has completed the computation for such theory, the box will be green. If the computation has been interrupted after six hours, the box is red. Grey boxes take into account the fact that the table are symmetric, since $(G,G')\sim (G',G)$. For the green boxes we claim such theories are either in the list (\ref{list:knownlag}) or if not then a UV lagrangian cannot exist. For red boxes we ignore if a Maruyoshi-Song UV lagrangian can exist or not. In Table~\ref{tab:percentage} we also list the percentage of completeness of the computation for all the combinations of $(G,G')$ theories that have been analyzed.\\

Given the large set of $(G,G')$ theories covered in this scan, and given that it was possible to prove that all the analyzed cases do not admit a Maruyoshi-Song lagrangian, we are lead to make the following conjecture:\\

\noindent\fbox{%
    \parbox{\textwidth}{
\textit{\noindent Let $\mathcal{T}_{\IR}$ be any theory of $(G,G')$ type. Then either a Maruyoshi-Song lagrangian $\mathcal{T}_{\UV}$ flowing to $\mathcal{T}_{\IR}$ is already known, or if not, then it does not exist.
}
}%
}\\

We stress here again the fact that we are only conjecturing non-existence of UV lagrangians for certain theories of $(G,G')$ type, and only under the hypothesis that the UV theory flows to $(G,G')$ under a Maruyoshi-Song mechanism. Of course the $(G,G')$ landscape does not exhaust all the possible Argyres-Douglas theories, and of course there could be other methods, different from the Maruyoshi-Song deformation, with which a UV lagrangian can flow to an Argyres-Douglas theory, regardless if it is of $(G,G')$ type or not. However, our program is not only restricted to theories of $(G,G')$ type: it can test the existence of a Maruyoshi-Song lagrangian, for any theory $\mathcal{T}_{\IR}$ with known central charges and rank. It would be interesting to investigate more on such possibilities.

For a better visualization of the results, the tables for the theories $(A_n,A_m)$, $(A_n,D_m)$ and $(D_n,D_m)$ have been split in four tables each. For a $(G,G')$ theory, we show the group $G$ in the rows and the group $G'$ in the columns. In Tables from~\ref{tab:AnAm1} to~\ref{tab:AnAm4} there are the $(A_n,A_m)$ theories. The $(G,G')$ theories, in this case, are symmetrical under the permutation of $(G,G')$. The gray boxes are associated to theories already shown in the corresponding symmetrical case. In Tables from~\ref{tab:DnAm1} to~\ref{tab:DnAm4} there are the $(A_n,D_m)$ theories. Here, the green boxes are associated to the combinations involving $D_1$ or $D_2$ which are not computed by the program. In Tables from~\ref{tab:DnDm1} to~\ref{tab:DnDm4} there are the $(D_n,D_m)$ theories. Here, again, we have gray boxes which are associated to theories already shown in the symmetrical case and to the combinations involving $D_1$ or $D_2$. In Table~\ref{tab:AnDmE678} there are the results for the theories $(A_n,E_6)$, $(A_n,E_7)$, $(A_n,E_8)$ with $n=1,\ldots, 100$ and $(D_m,E_6)$, $(D_m,E_7)$, $(D_m,E_8)$ with $m=3,\ldots,100$. Finally, for completeness, we also show the table for $(G,G')$ theories where $G$ and $G'$ are $E_6$, $E_7$ and $E_8$ in Table~\ref{tab:EnEm}. 

The scan has been carried out on the IFT \emph{Hydra cluster} and the DESY \emph{theoc cluster}, in parallel computation on an average of $9$ cores for $1$ month, and then on the same cluster with an average of $16$ cores for $2$ months. The total single-core CPU time is then approximately 41 months. The CPU are Intel(R) Xeon(R) CPU E5-2650 v2 @ 2.60GHz.

\begin{table}[h!]
    \centering
    \begin{tabular}{c|c}
      $(G,G')$   & Analyzed theories \\
      \hline
       $(A_n,A_m)$  & $82,85\%$\\
        \hline
       $(A_n,D_m)$  & $78,70\%$\\
        \hline
       $(D_n,D_m)$  & $77,42\%$\\
        \hline
       $(A_n,E_6)$  & $86,00\%$\\
        \hline
       $(A_n,E_7)$  & $84,00\%$\\
        \hline
       $(A_n,E_8)$  & $82,00\%$\\
        \hline
       $(D_m,E_6)$  & $88,66\%$\\
        \hline
       $(D_m,E_7)$  & $80,41\%$\\
        \hline
       $(D_m,E_8)$  & $81,44\%$\\
       \hline 
       $(E_i,E_j)$ & $100,00\%$\\
       \hline 
       Total & $79,60\%$
    \end{tabular}
    \caption{Percentage of theories for which the computation has terminated without being interrupted after six hours. The subscripts $i,j=6,7,8$.}
    \label{tab:percentage}
\end{table}

\begin{table}[h!]
    \centering
    \includegraphics[width=\textwidth,height=\textheight]{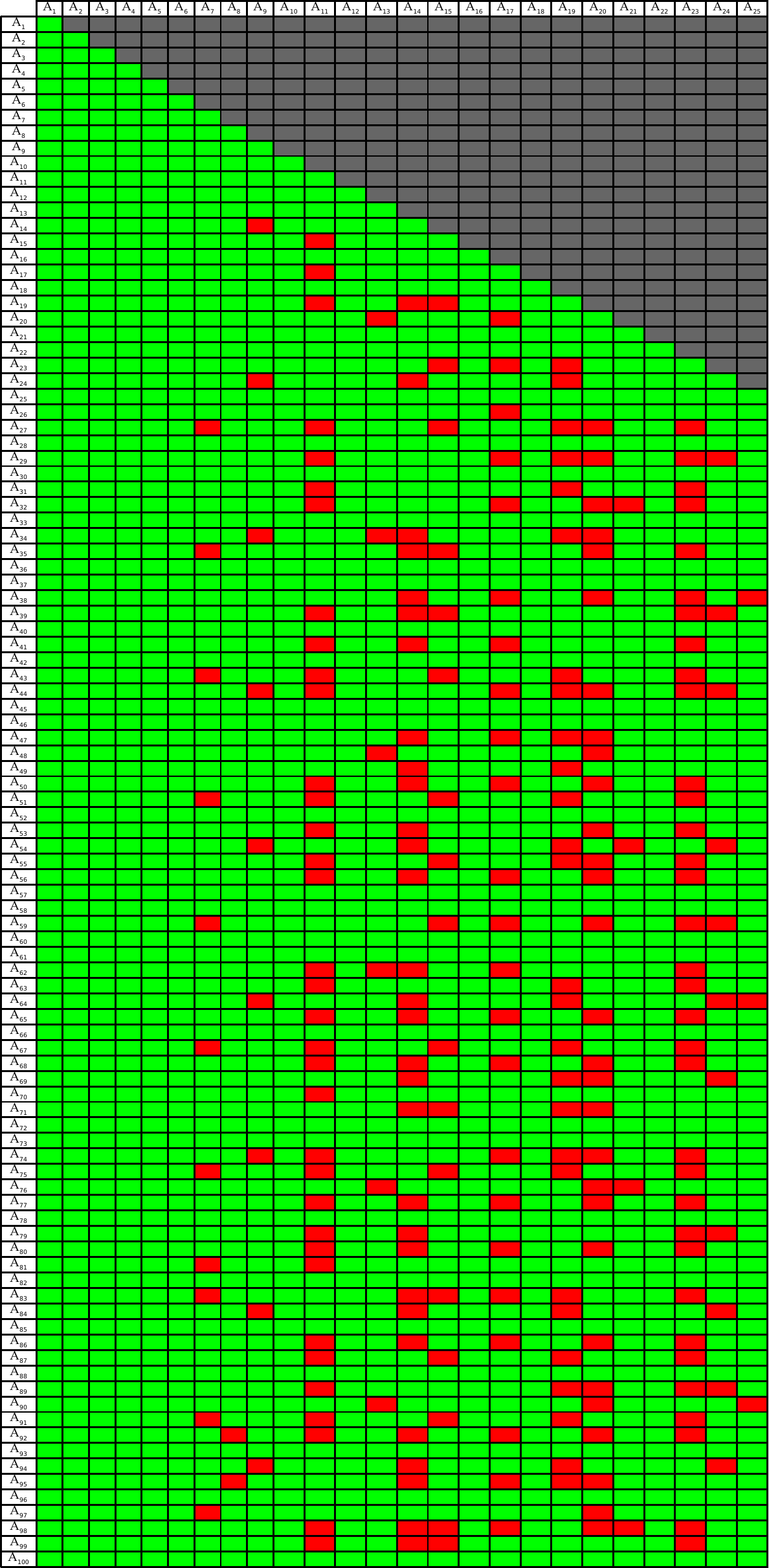}
    \caption{Analyzed theories for $(A_n,A_m)$ with $n=1,\ldots 100$ and  $m=1,\ldots 25$.}
    \label{tab:AnAm1}
\end{table}

\begin{table}[h!]
    \centering
    \includegraphics[width=\textwidth,height=\textheight]{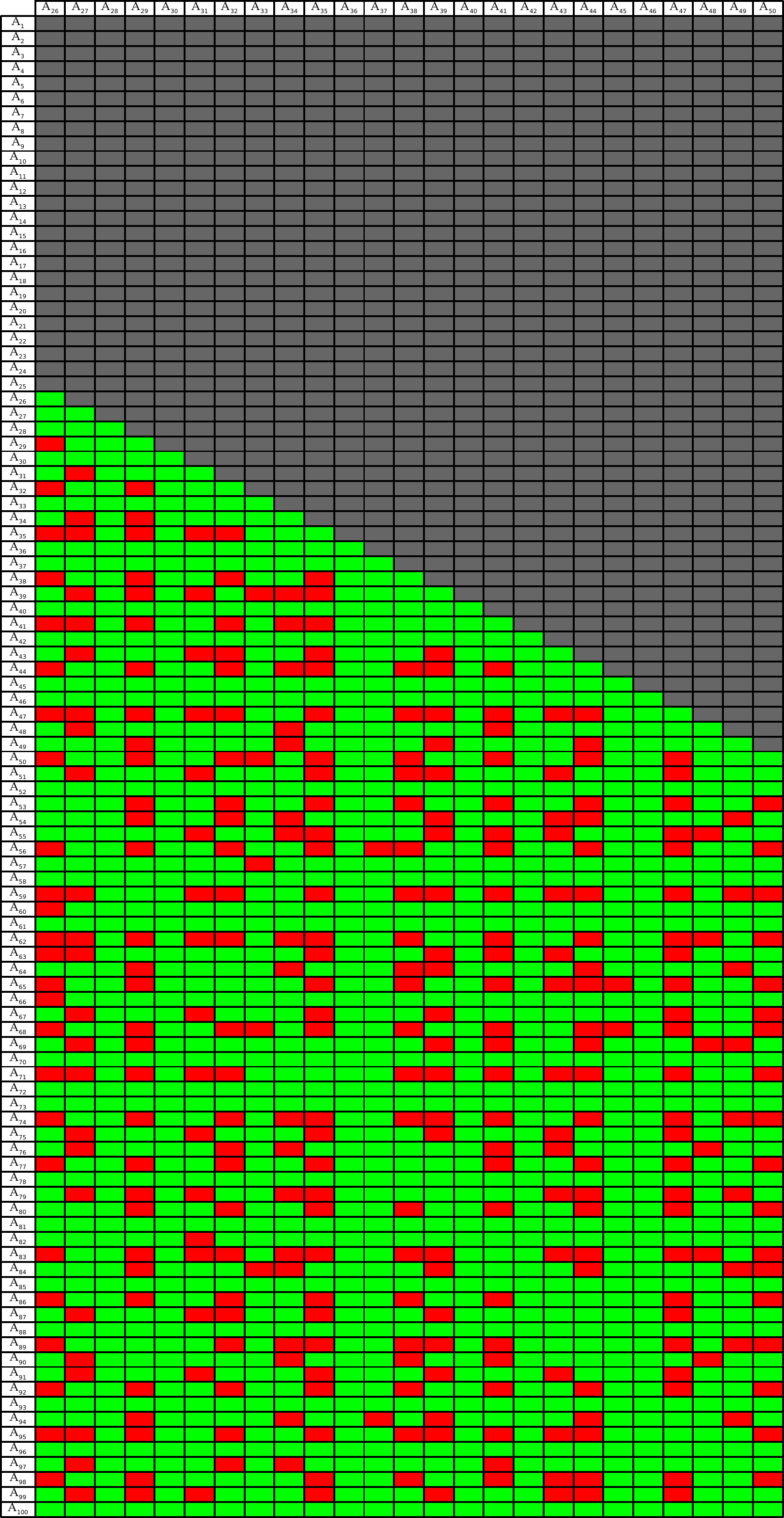}
    \caption{Analyzed theories for $(A_n,A_m)$ with $n=1,\ldots 100$ and  $m=26,\ldots 50$.}
    \label{tab:AnAm2}
\end{table}

\begin{table}[h!]
    \centering
    \includegraphics[width=\textwidth,height=\textheight]{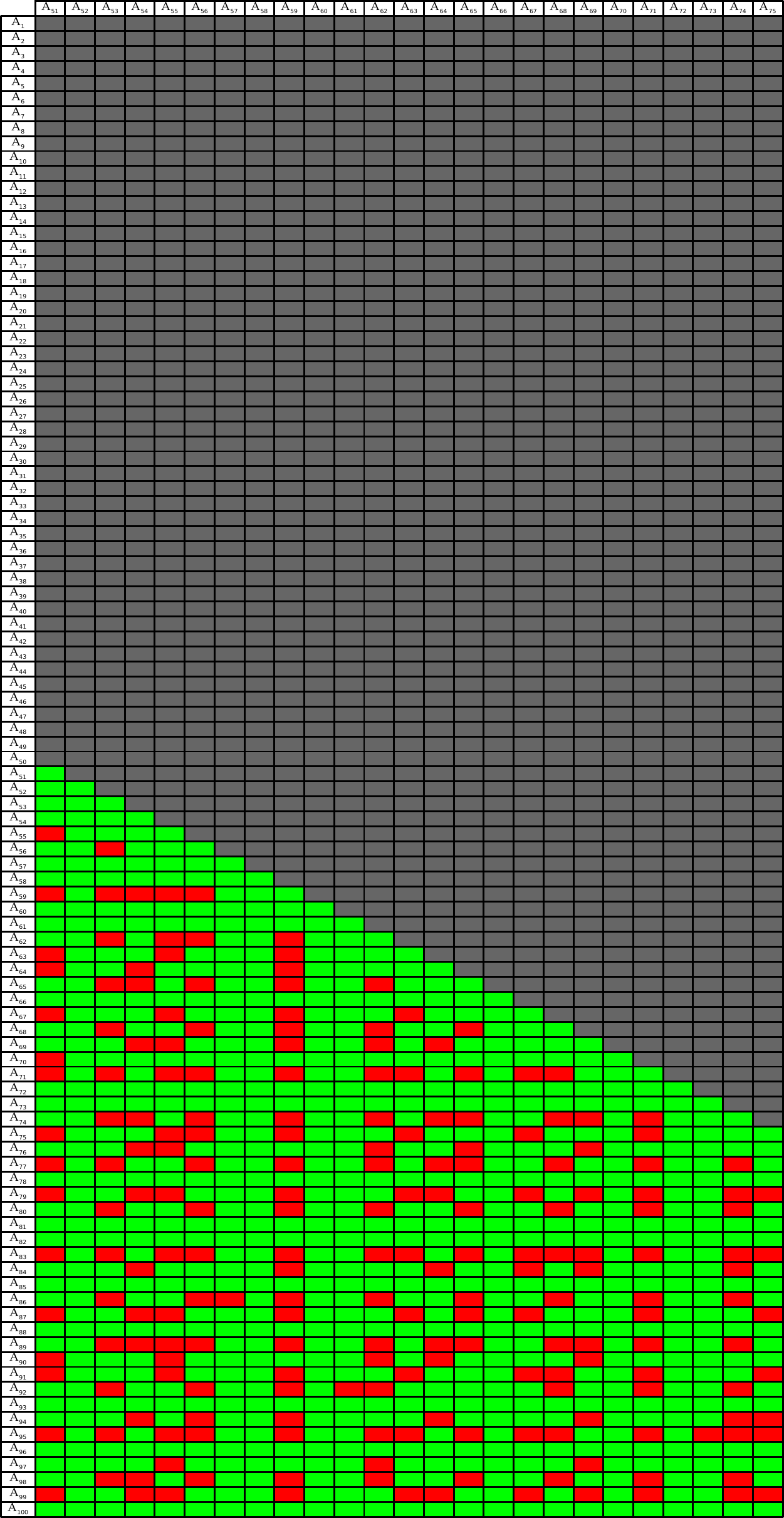}
    \caption{Analyzed theories for $(A_n,A_m)$ with $n=1,\ldots 100$ and  $m=51,\ldots 75$.}
    \label{tab:AnAm3}
\end{table}

\begin{table}[h!]
    \centering
    \includegraphics[width=\textwidth,height=\textheight]{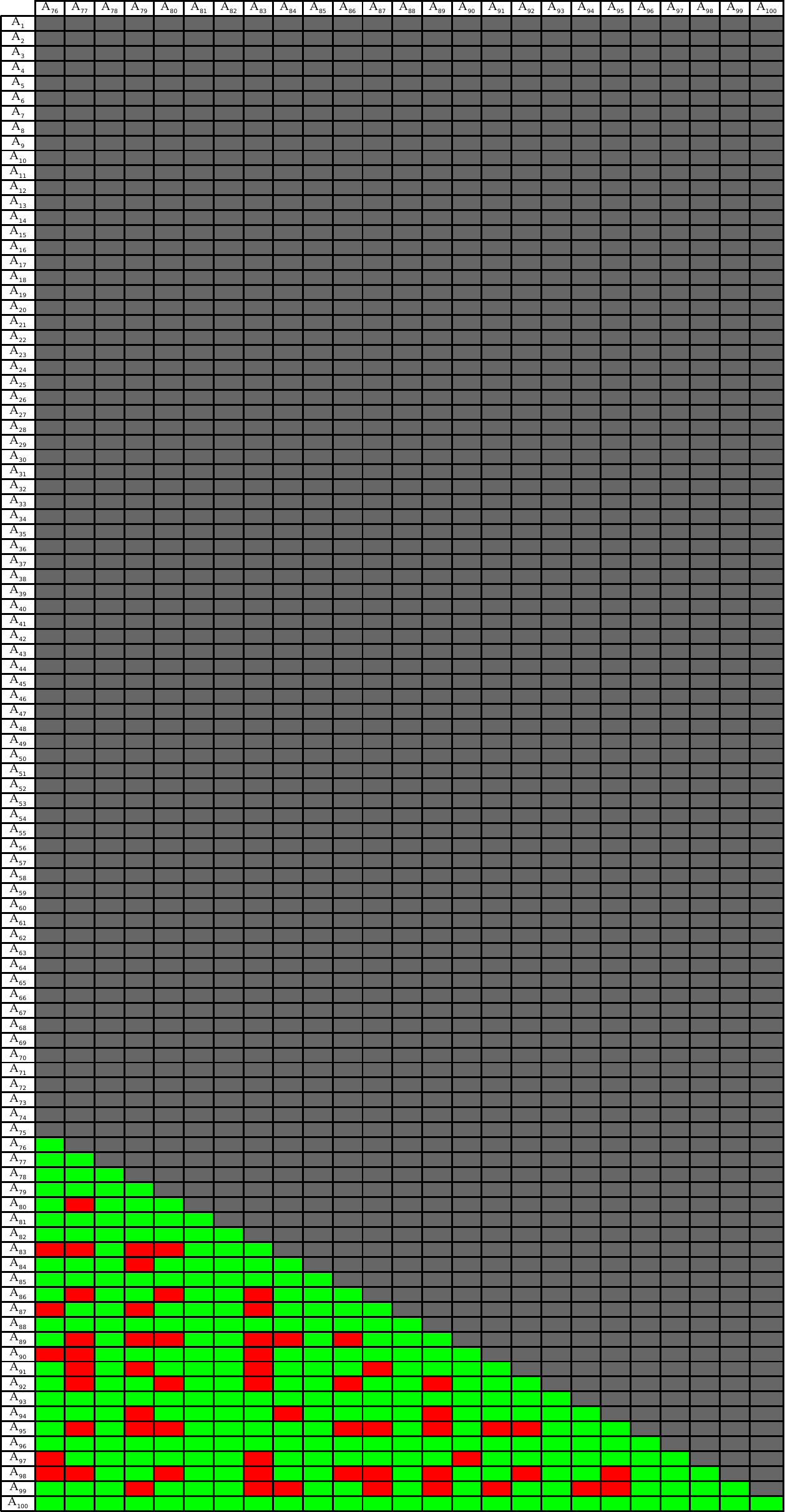}
    \caption{Analyzed theories for $(A_n,A_m)$ with $n=1,\ldots 100$ and  $m=75,\ldots 100$.}
    \label{tab:AnAm4}
\end{table}

\begin{table}[h!]
    \centering
    \includegraphics[width=\textwidth,height=\textheight]{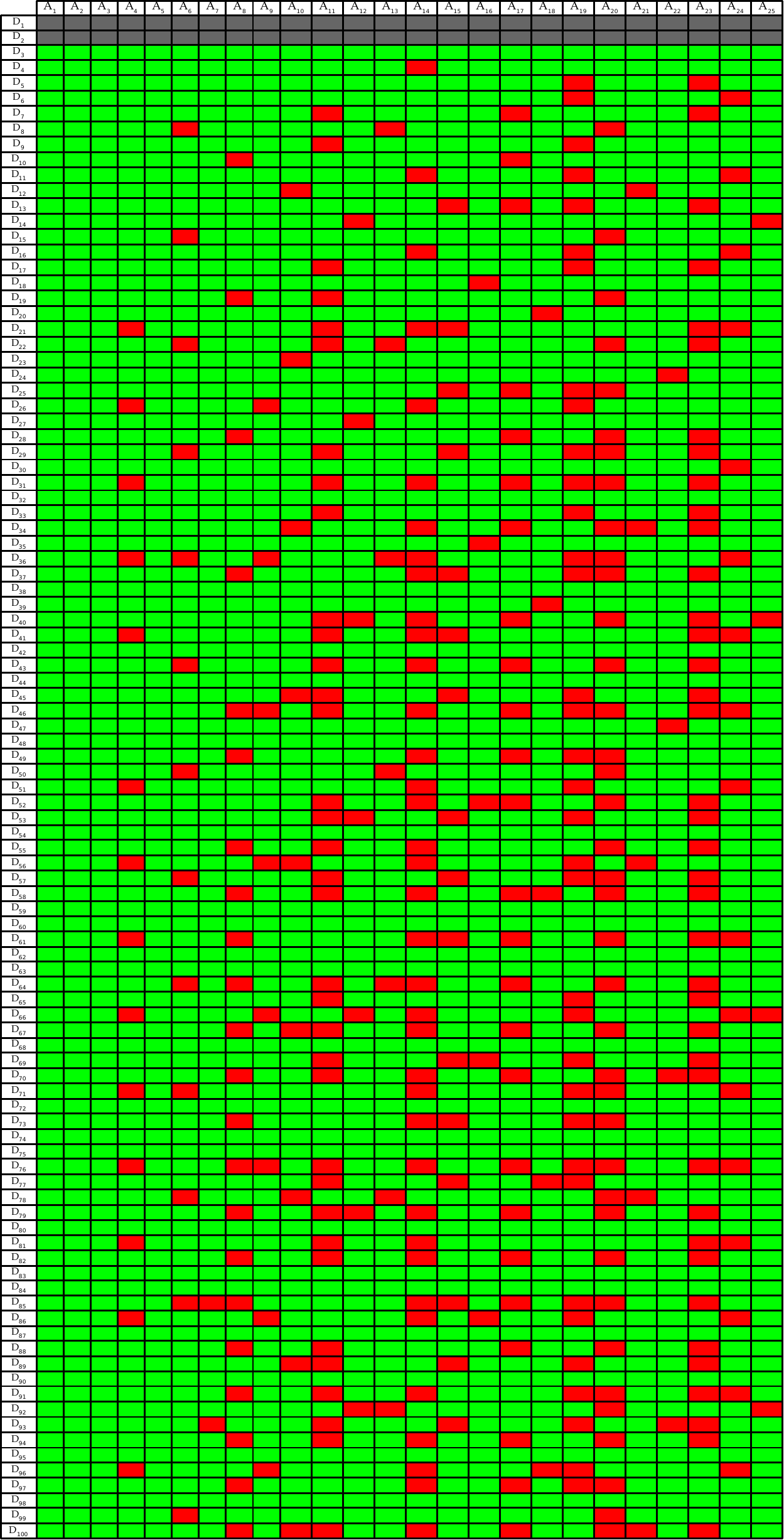}
    \caption{Analyzed theories for $(D_n,A_m)$ with $n=3,\ldots 100$ and  $m=1,\ldots 25$.}
    \label{tab:DnAm1}
\end{table}

\begin{table}[h!]
    \centering
    \includegraphics[width=\textwidth,height=\textheight]{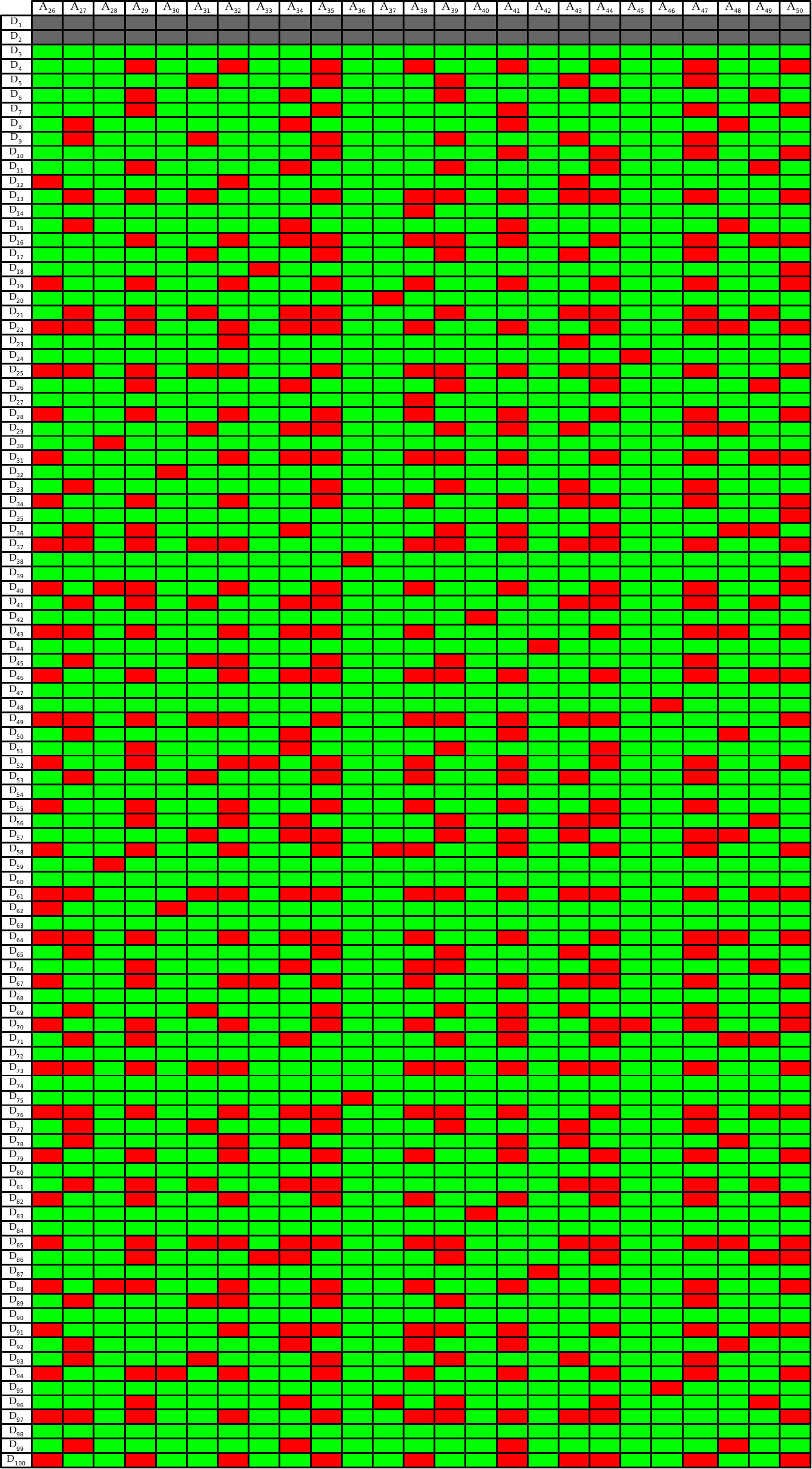}
    \caption{Analyzed theories for $(D_n,A_m)$ with $n=3,\ldots 100$ and  $m=26,\ldots 50$.}
    \label{tab:DnAm2}
\end{table}

\begin{table}[h!]
    \centering
    \includegraphics[width=\textwidth,height=\textheight]{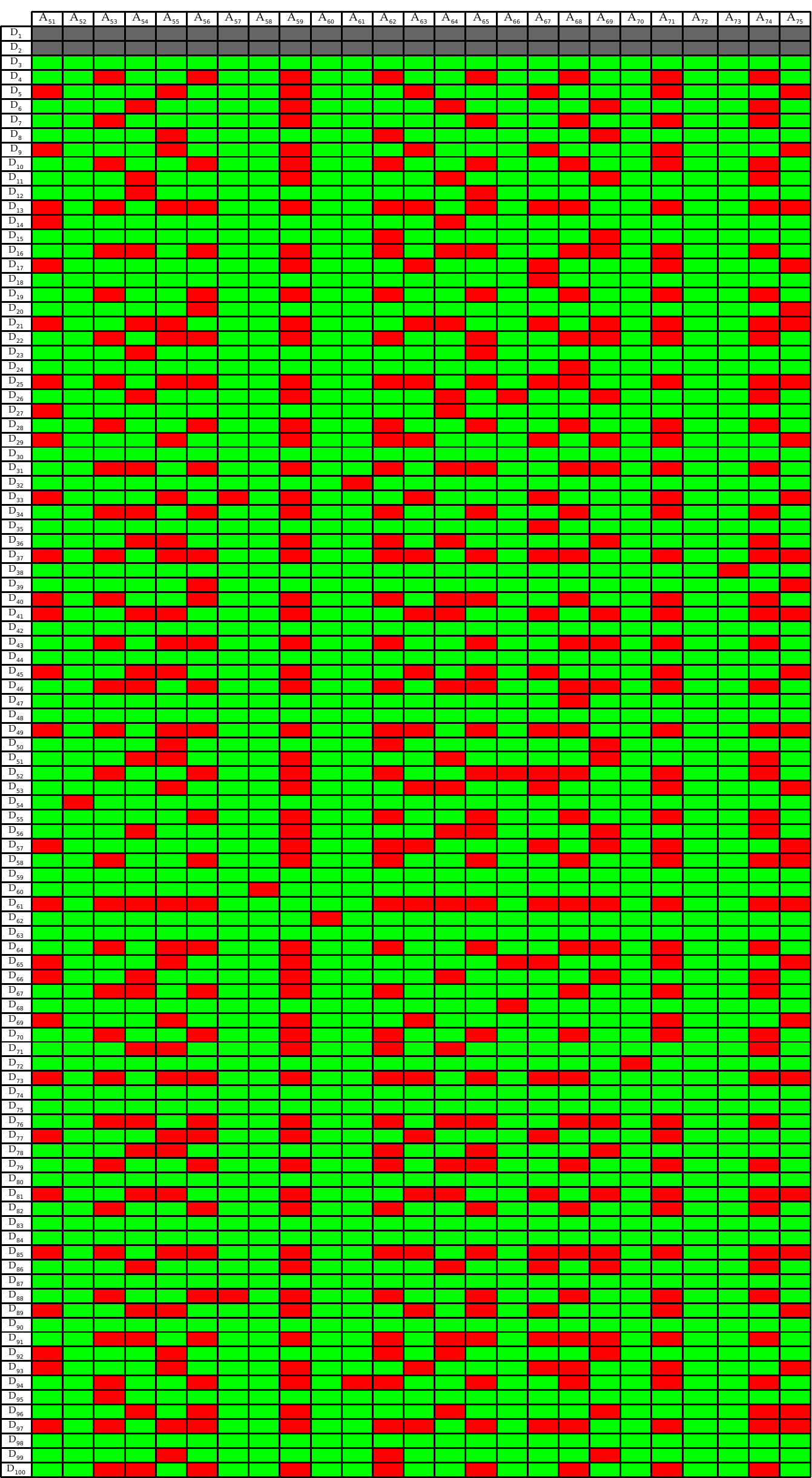}
    \caption{Analyzed theories for $(D_n,A_m)$ with $n=3,\ldots 100$ and  $m=51,\ldots 75$.}
    \label{tab:DnAm3}
\end{table}

\begin{table}[h!]
    \centering
    \includegraphics[width=\textwidth,height=\textheight]{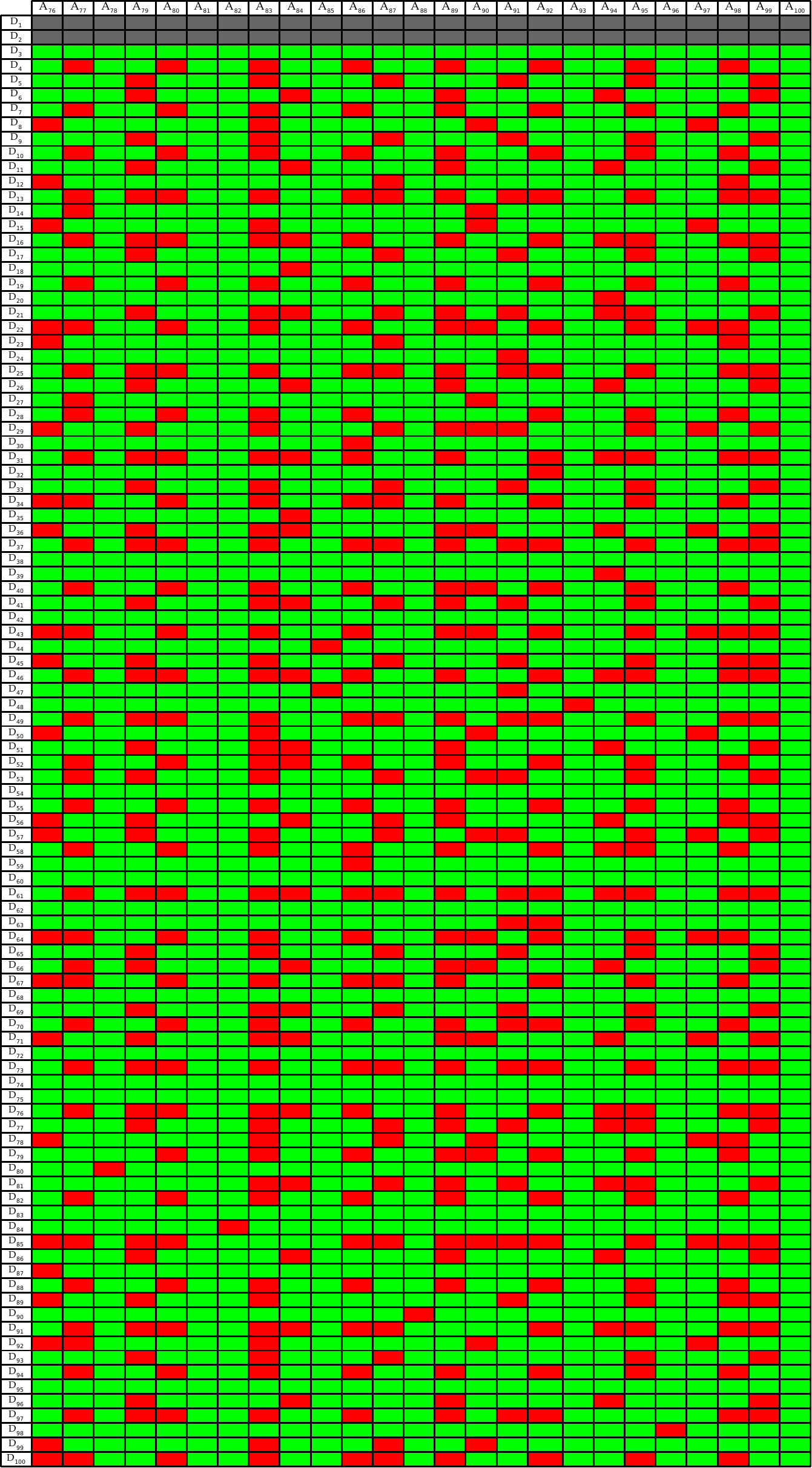}
    \caption{Analyzed theories for $(D_n,A_m)$ with $n=3,\ldots 100$ and  $m=75,\ldots 100$.}
    \label{tab:DnAm4}
\end{table}

\begin{table}[h!]
    \centering
    \includegraphics[width=\textwidth,height=\textheight]{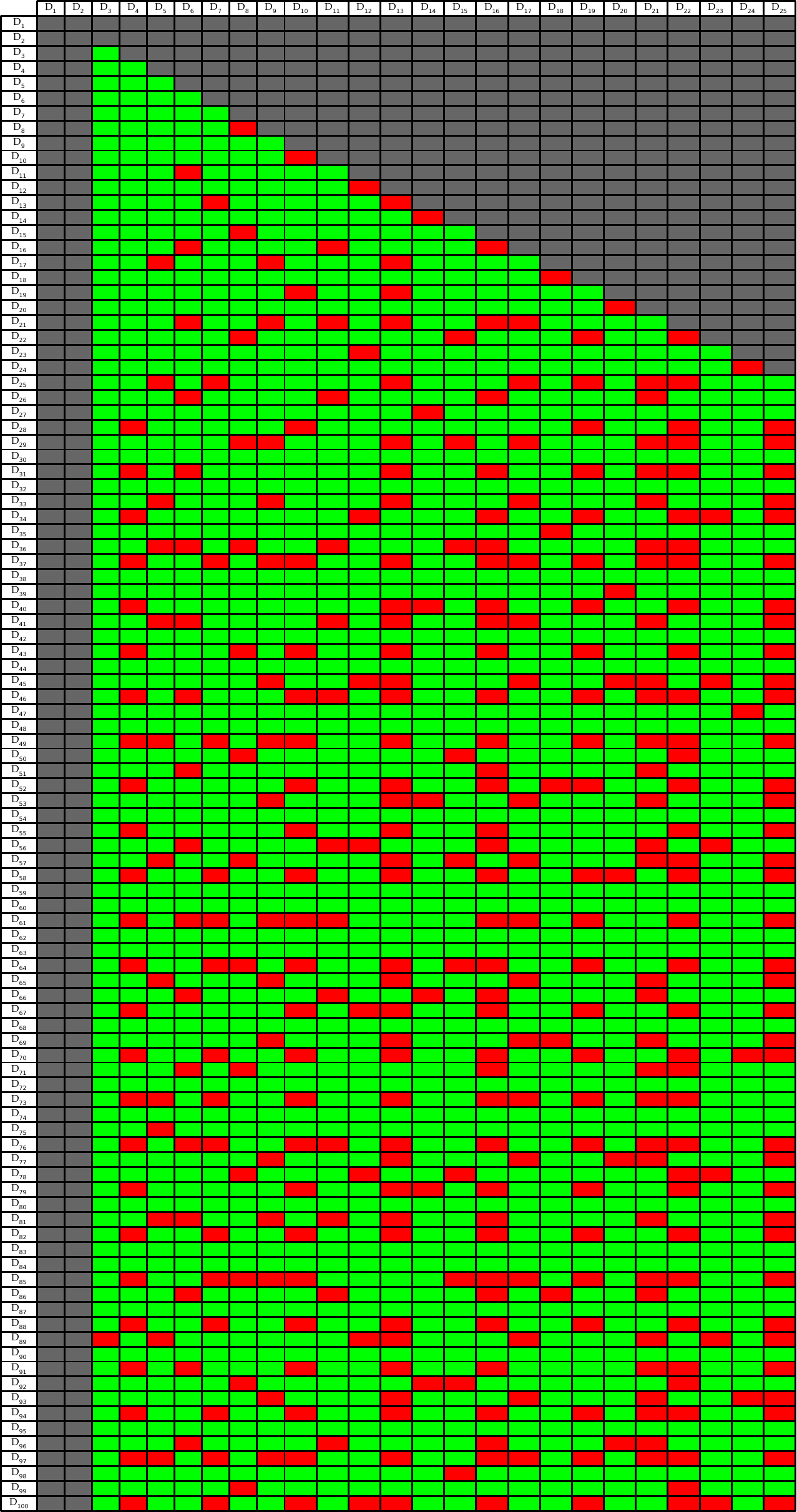}
    \caption{Analyzed theories for $(D_n,D_m)$ with $n=3,\ldots 100$ and  $m=3,\ldots 25$.}
    \label{tab:DnDm1}
\end{table}

\begin{table}[h!]
    \centering
    \includegraphics[width=\textwidth,height=\textheight]{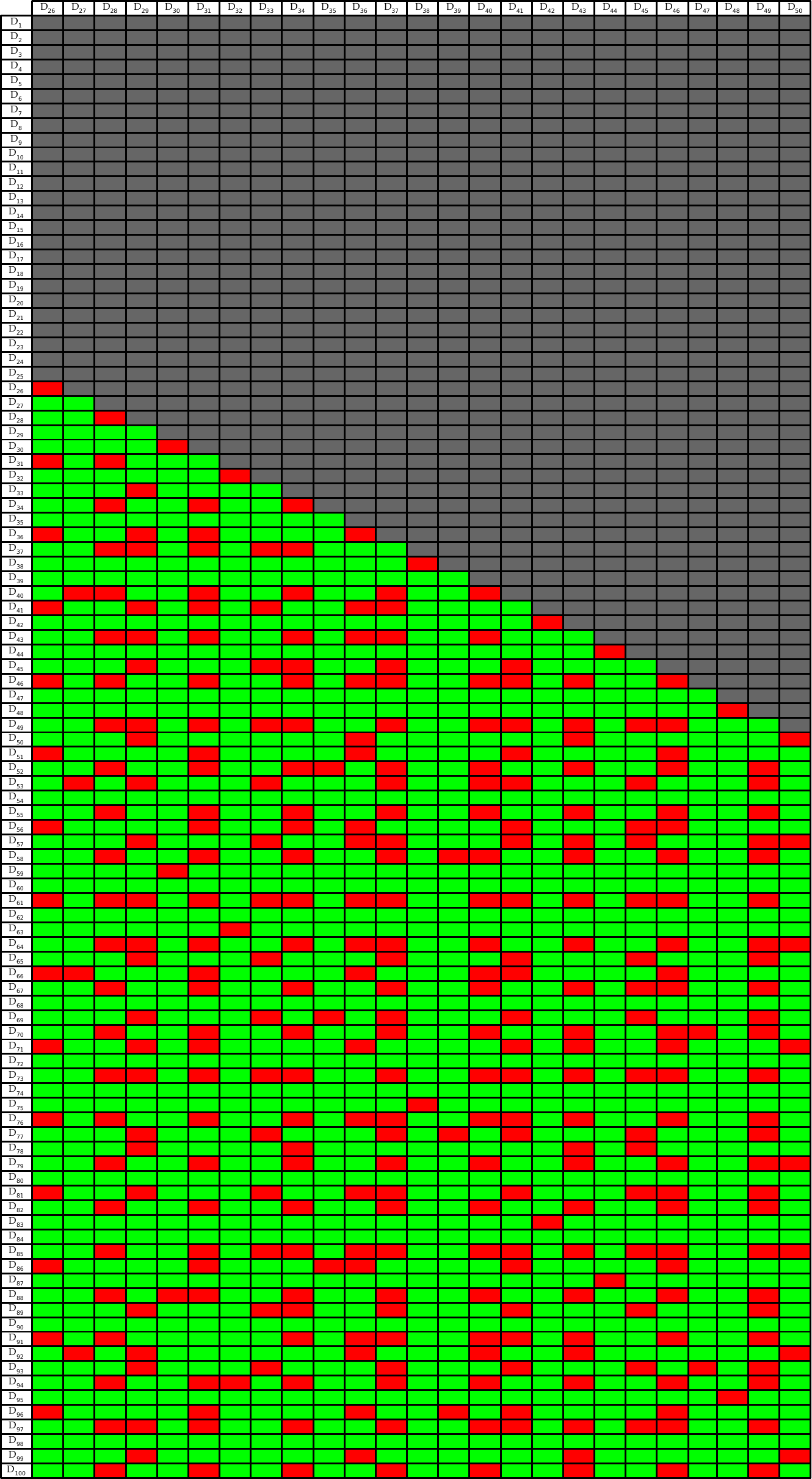}
    \caption{Analyzed theories for $(D_n,D_m)$ with $n=3,\ldots 100$ and  $m=26,\ldots 50$.}
    \label{tab:DnDm2}
\end{table}

\begin{table}[h!]
    \centering
    \includegraphics[width=\textwidth,height=\textheight]{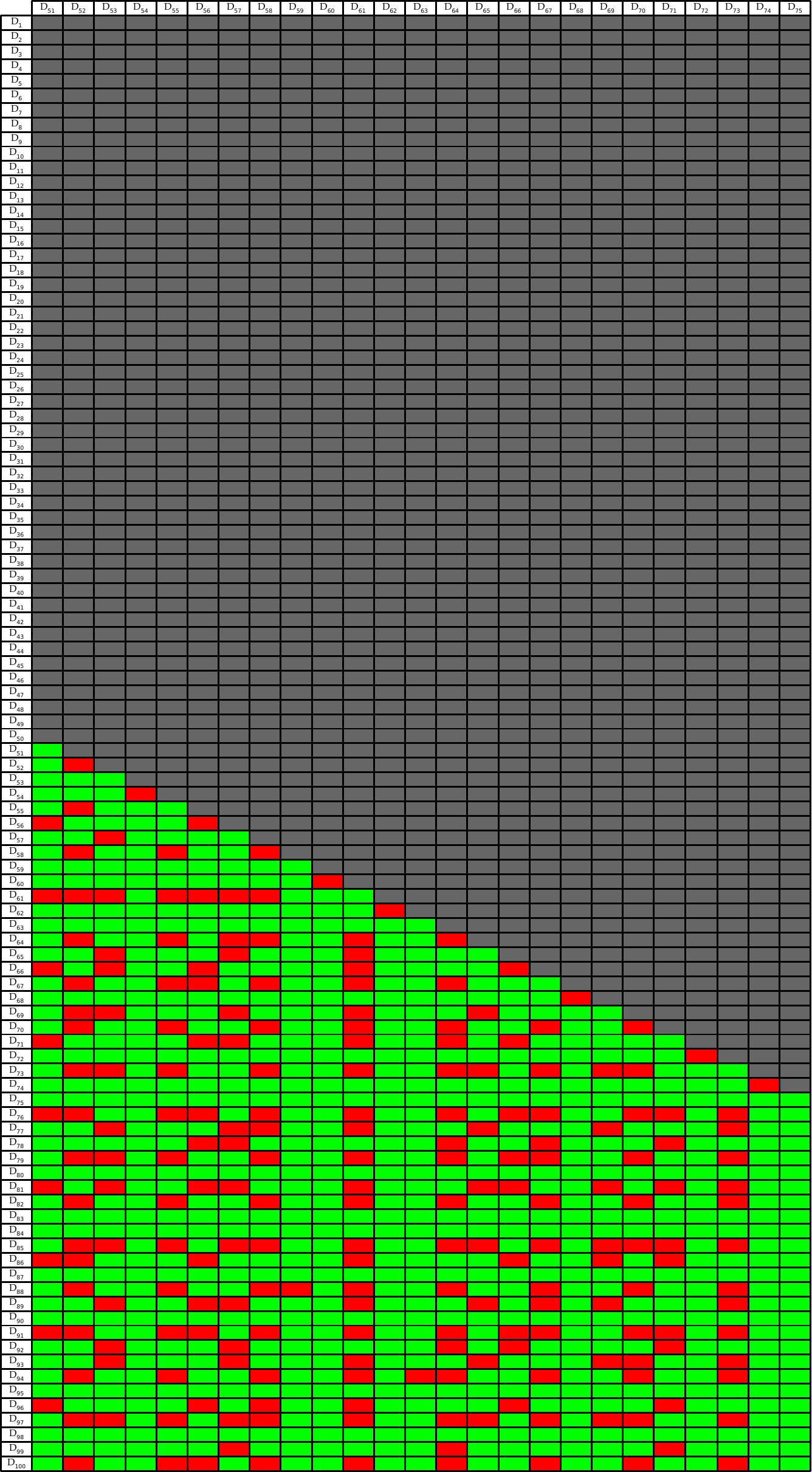}
    \caption{Analyzed theories for $(D_n,D_m)$ with $n=3,\ldots 100$ and  $m=51,\ldots 75$.}
    \label{tab:DnDm3}
\end{table}

\begin{table}[h!]
    \centering
    \includegraphics[width=\textwidth,height=\textheight]{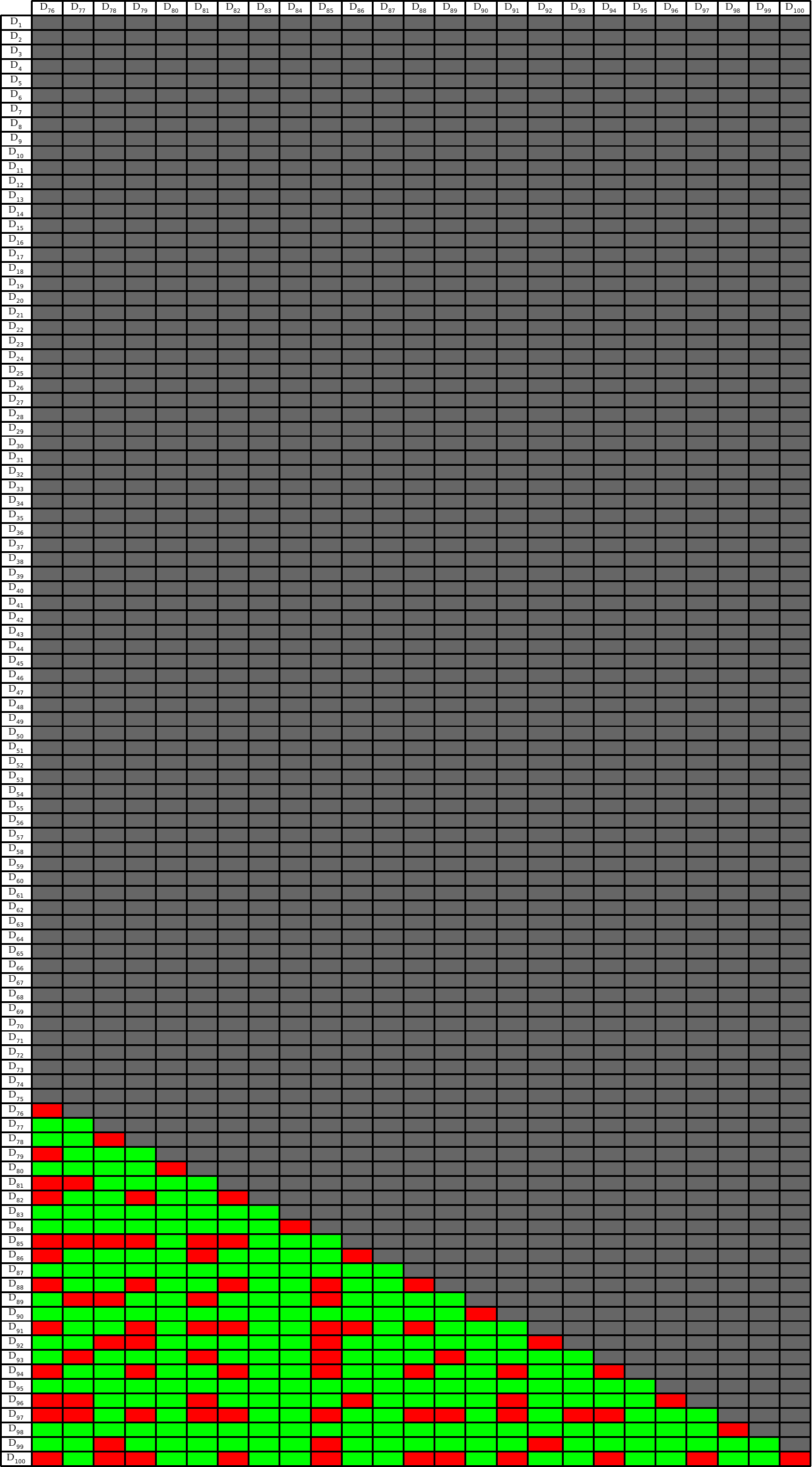}
    \caption{Analyzed theories for $(D_n,D_m)$ with $n=3,\ldots 100$ and  $m=76,\ldots 100$.}
    \label{tab:DnDm4}
\end{table}

\begin{table}[h!]
    \begin{minipage}{0.50\textwidth}
    \centering
    \includegraphics[height=\textheight,keepaspectratio]{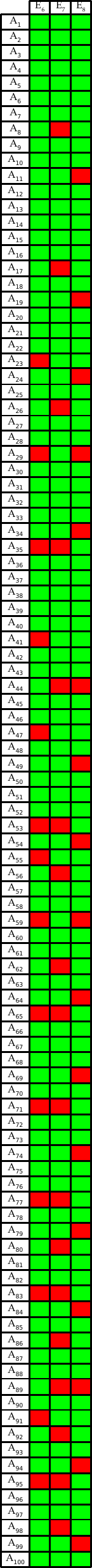}
    \end{minipage}
    \begin{minipage}{0.50\textwidth}
    \centering
    \includegraphics[height=\textheight,keepaspectratio]{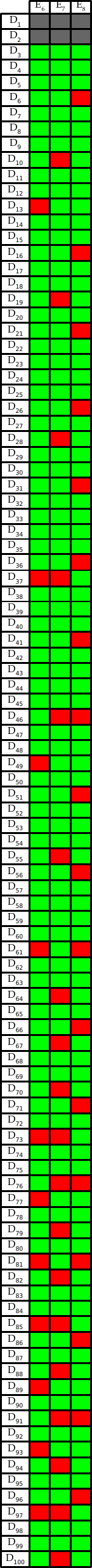}
    \end{minipage}
    \caption{Analyzed theories for $(A_n,E_6)$, $(A_n,E_7)$ and $(A_n,E_8)$ with $n=1,\ldots, 100$ and $(D_m,E_6)$, $(D_m,E_7)$ and $(D_m,E_8)$, with $n=3,\ldots 100$.}
    \label{tab:AnDmE678}
\end{table}

\begin{table}[h!]
\centering
    \includegraphics[width=0.25\textwidth,keepaspectratio]{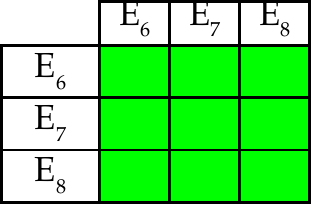}
\caption{Analyzed theories for all the combinations of $(G,G')$ given by the groups $E_6$, $E_7$ and $E_8$.}
\label{tab:EnEm}
\end{table}

\FloatBarrier


\bigskip

\centerline{\bf \large Acknowledgments}

FC would like to thank Simone Giacomelli, Alessandro Pini and Raffaele Savelli for discussions and comments on the draft. The work of FC is supported by the ERC Consolidator Grant STRINGFLATION under the HORIZON 2020 grant agreement no. 647995.
AM would like to thank Florent Baume, Emilio Ambite and Marcos Ram\'irez for the support with the HYDRA cluster in the IFT. AM received funding from ``la Caixa" Foundation (ID 100010434) with fellowship code LCF/BQ/IN18/11660045 and from the European Union’s Horizon 2020 research and innovation programme under the Marie Sk\l odowska-Curie grant agreement No. 713673.

\bibliography{RefsGgpScan.bib}
\bibliographystyle{JHEP}

\end{document}